\documentclass{Interspeech2023}

\interspeechcameraready

\usepackage{amsmath,graphicx}

\usepackage[utf8]{inputenc} % allow utf-8 input
\usepackage[T1]{fontenc}    % use 8-bit T1 fonts
\usepackage{hyperref}       % hyperlinks
\usepackage{url}            % simple URL typesetting
\usepackage{booktabs}       % professional-quality tables
\usepackage{amsfonts}       % blackboard math symbols
\usepackage{nicefrac}       % compact symbols for 1/2, etc.
\usepackage{microtype}      % microtypography
\usepackage{xcolor}         % colors
\usepackage{graphicx}% so it makes black blobs
\usepackage{multicol}
\usepackage{multirow}
\usepackage{color, colortbl}
\usepackage{amsbsy}
\usepackage{array, makecell} %
\usepackage{boldline} 
\usepackage{hhline}
\usepackage[numbib]{tocbibind}
\usepackage[numbers,sort&compress]{natbib}
\usepackage[labelfont=bf]{caption}
\usepackage{todonotes}
\newcolumntype{P}[1]{>{\centering\arraybackslash}p{#1}}
\newcolumntype{M}[1]{>{\centering\arraybackslash}m{#1}}
\newcolumntype{N}{@{}m{0pt}@{}}
\definecolor{LightGray}{gray}{0.9}

\title{AG-LSEC: Audio Grounded Lexical Speaker Error Correction}
\name{Rohit Paturi, Xiang Li, Sundararajan Srinivasan}
\address{AWS AI Labs}
\email{paturi@amazon.com, xiangzai@amazon.com, sundarsr@amazon.com}
\begin{document}
\maketitle
\begin{abstract}
Speaker Diarization (SD) systems are typically audio-based and operate independently of the ASR system in traditional speech transcription pipelines and can have speaker errors due to SD and/or ASR reconciliation, especially around speaker turns and regions of speech overlap. To reduce these errors, a Lexical Speaker Error Correction (LSEC), in which an external language model provides lexical information to correct the speaker errors, was recently proposed. Though the approach achieves good Word Diarization error rate (WDER) improvements, it does not use any additional acoustic information and is prone to miscorrections. In this paper, we propose to enhance and acoustically ground the LSEC system with speaker scores directly derived from the existing SD pipeline. This approach achieves significant relative WDER reductions in the range of 25-40\% over the audio-based SD, ASR system and beats the LSEC system by 15-25\% relative on RT03-CTS, Callhome American English and Fisher datasets.
%without needing any additional model parameters.

\end{abstract}

\noindent\textbf{Index Terms}: Speaker Diarization, Language Models, Automatic Speech Recognition, Error Correction
\section{Introduction}

Multi-talker speech transcription systems answer the question “Who spoke what and when?” in an audio recording. Though single-talker speech transcription systems have advanced significantly in the recent years, there are still challenges in transcribing and diarizing multi-talker speech due to the regular speaker changes and overlapping speech in natural conversations. Multi-talker transcription systems can be broadly classified in three categories of systems that aim to transcribe and diarize multi-talker speech: 1) Speech/Speaker Separation followed by ASR \cite{von2020end,9413423,paturi22_interspeech,zhao2023mossformer}, 2) Speaker attributed ASR \cite{chang2020end, kanda22_interspeech,shi23d_interspeech,sklyar2021streaming, berger23_interspeech} and 3) Modular ASR \cite{wang2020transformer,yang2022conformer, rekesh2023fast, radford2023robust}, SD systems \cite{park2022review,wang2018speaker,fujita2019end,kinoshita2021integrating, bredin23_interspeech}. 

Speech Separation systems separate speech from multiple speakers into different channels while performing ASR individually on these channels, thus assigning a speaker label to each transcribed word. These systems are usually trained with Permutation Invariant Training (PIT) which constrains the maximum number of speakers it can handle \cite{von2020end,9413423}, and is also prone to duplicated artifacts across channels among other challenges to handle long-form audio \cite{paturi22_interspeech}. Speaker attributed ASR (SA-ASR) systems either follow PIT based training \cite{sklyar2021streaming, berger23_interspeech} which have the same challenges as above or the more recent Serialized Output Training (SOT) \cite{kanda22_interspeech,shi23d_interspeech}  systems which outputs transcripts of each speaker sequentially. Main drawbacks with SOT systems are the lack of timestamps for the speaker turns needed for downstream applications, and the difficulty of speaker reconciliation across transcribed blocks for long-form audios.  Training both speech separation and SA-ASR systems is data intensive and needs large amounts of multi-channel audio data. Such data is difficult to source, especially for domains outside of telephony, and could also increase cost significantly. Finally, independent modular ASR and SD systems is one of the most successful and practical way to transcribe multi-talker audios. The ASR and SD systems operate independently and their results are reconciled to assign speaker labels to each transcribed word \cite{shafey2019joint}. Such systems can naturally transcribe and diarize long-form audio by operating on independent speech segments without significantly constraining the maximum number of speakers, though these approaches are error-prone around speaker change or overlapping speech regions due to SD or reconciliation errors. 

Most SD systems in literature rely only on acoustic information \cite{wang2018speaker,fujita2019end} to either extract speaker embeddings \cite{wan2018generalized, dawalatabad2021ecapa} followed by clustering the embeddings or directly perform diarization with End-to-End Neural Diarization (EEND) \cite{fujita2019end, kinoshita2021integrating}. Reliance on only acoustic information can lead to speaker errors, mainly around the speaker turns and regions of speech overlaps \cite{park2020speaker, xia2022turn}. In addition to the SD errors, speakers can be attributed to the wrong words in the SD-ASR reconciliation phase also due to errors solely in ASR word timings \cite{paturi23_interspeech}. The reconciliation errors will be higher in regions of speech overlap as a traditional ASR system can identify words corresponding to one of the overlapping speakers which can be wrongly attributed to the other overlapping speaker. 

% \begin{figure}[t]
%   \centering
%   \includegraphics[width=0.7\linewidth]{ASR_timing_errors_v2}
%   \caption{Errors in reconciliation of independent SD and ASR systems due to inaccurate ASR timings}
%   \vspace{-1\baselineskip}
% \end{figure}
\begin{figure*}[htp]
  \centering
  \includegraphics[width=0.9\linewidth]{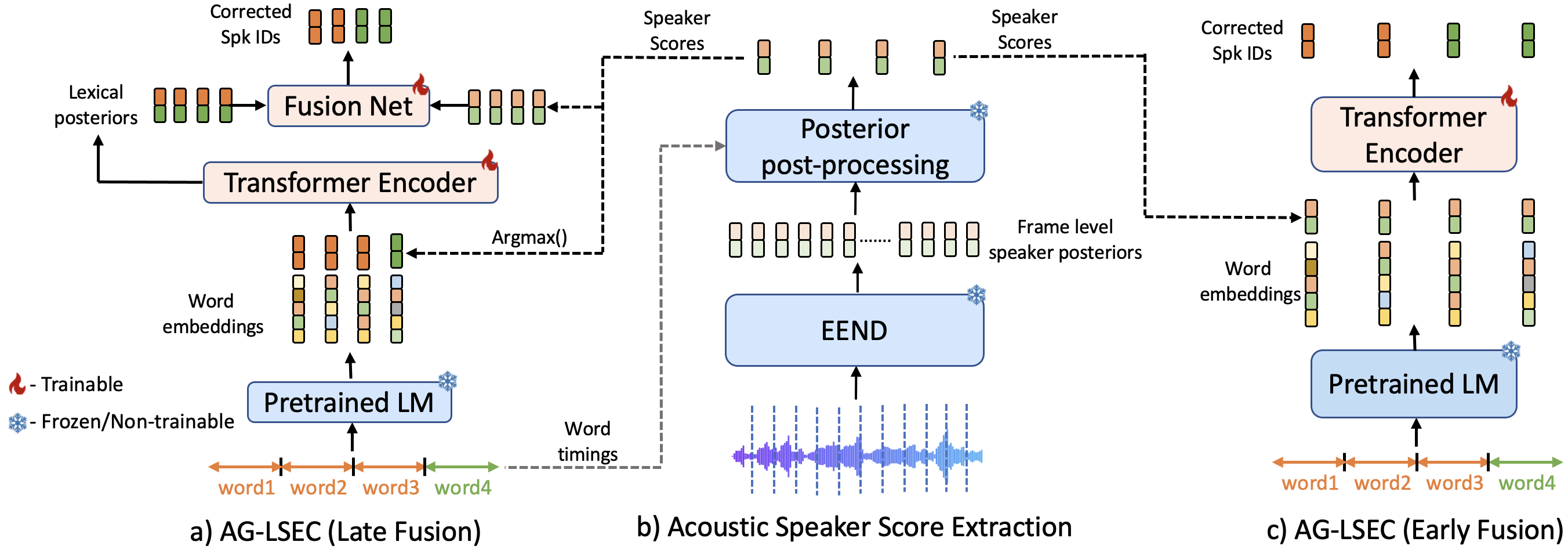}
  \vspace{-0.5\baselineskip}
  \caption{(a) AG-LSEC with Late Fusion of Speaker Scores, (b) Speaker Scores Extraction form the SD pipeline, (c) AG-LSEC with Early Fusion of Speaker Scores}
  \vspace{-1\baselineskip}
\end{figure*}

Lexical information often contains complementary information which can be useful in accurately assigning speaker labels to the transcribed words \cite{park2020speaker, xia2022turn, shafey2019joint}. For instance, analyzing only the written transcript of a conversation, "how are you i am good", enables us to infer that there is likely a speaker change between the utterances "how are you" and "i am good". There have been few works \cite{park2020speaker, xia2022turn} which leverage lexical information to perform standalone SD. Though these reduce the SD errors, these are still prone to ASR reconciliation errors around speaker turns and overlapped speech similar to the other acoustic only approaches. Another work \cite{shafey2019joint} models ASR and SD jointly while \cite{adedeji2024sound} employs a large language model (LLM)  to assign speaker roles
to a non-speaker-attributed transcript, but both these works are limited to two speakers with distinct roles.
%Also, both these works need ASR before performing diarization thus increasing the latency of the system. 
\section{Related Work}

Recent works \cite{wang2024diarizationlm, park2023enhancing} have explored leveraging a LLM for performing word level SD or correcting speaker errors after ASR reconciliation. \cite{park2023enhancing} models SD as a beam search at the word level, with distinct speaker paths for each word in the transcript, where the probability that a given word is emitted by a particular speaker is computed using a LLM. However, a significant drawback of this approach is the high computational cost and latency involved in prompting the LLM for each word in the transcript, as well as the need to tune the LLM to emit the response in a specific format, which otherwise can be susceptible to the non-deterministic nature of the LLM.

%One approach \cite{adedeji2024sound} employs an LLM to assign speaker roles to a non-speaker-attributed transcript, but is limited to two speakers with distinct roles. 
The authors of \cite{wang2024diarizationlm} utilize a LLM to correct the outputs of a speaker-attributed transcript using a transcript preserving mechanism to prevent LLM hallucinations. They demonstrate that a zero or one-shot approach using a frozen LLM degrades diarization performance, and fine-tuning the LLM provides good WDER improvements. However, fine-tuning LLMs \cite{park2023enhancing, wang2024diarizationlm} can be data-intensive and computationally expensive and can result in higher inference latency. In addition, methods involving LLMs would warrant special mechanisms to prevent LLM hallucinations \cite{wang2024diarizationlm}.

An alternative method \cite{paturi23_interspeech} proposes a simple yet effective way to perform speaker error correction by applying a smaller encoder language model to the transcriptions and fusing these with speaker labels from SD to correct speaker and reconciliation errors. One limitation of \cite{paturi23_interspeech, wang2024diarizationlm, adedeji2024sound} is their reliance solely on lexical information, which can lead to both under-corrections and over-corrections, as conversational speech often contains artifacts such as speech overlaps, disfluencies, incomplete speaker turns, etc may not be captured in text-based language models. 

In this work, we propose to enhance the LSEC model in \cite{paturi23_interspeech} by grounding the model with acoustic information extracted from the speaker diarization module.
\section{Audio Grounded Lexical Error Corrector}
In this section, we introduce our proposed approach for leveraging both acoustic and lexical information from the speech signals to perform the speaker error correction. The proposed AG-LSEC consists of three main components:  an acoustic speaker score extractor, backbone language model (LM) and a Transformer Encoder Front-end to predict the speaker labels. We first introduce the extraction of speaker scores from the EEND system followed by two fusion techniques to incorporate the speaker scores into the lexical speaker error correction model.

\begin{table*}[!ht]
\caption{WDER of different models on Fisher test, RT03-CTS and CHAE sets.}
\vspace{-0.2\baselineskip}
\label{tab:data sampling table}
\centering
 \begin{tabular}{P{50mm}P{12mm}P{10mm}P{10mm}P{10mm}P{10mm}P{15mm}} 
 \hline\hline
     \multirow{2}{*}{Model Type}   &\multirow{2}{*}{\makecell{Fisher\\Test}}  & \multicolumn{2}{c}{RT03-CTS} & \multicolumn{2}{c}{CHAE} & \multirow{2}{*}{\makecell{CH109}} \\ \cline{3-6}
     & & Validation & Test & Validation & Test \\ \hline\hline
     Baseline (No Correction)  & 2.56  & 2.37 & 2.64 & 3.98 & 3.45 & 4.06 \\
     \makecell{LSEC \cite{paturi23_interspeech}} & 2.03 & 1.93 & 2.10 & 3.5 & 2.89 & 3.56 \\
     \makecell{AG-LSEC Late Fusion} & 1.80  & 1.70 & 1.80 & 3.26 & 2.72 & 3.32 \\
     % T2P1(P2): T2 init + Paired Data Tuning (P2) &  &  &  & 4.64 & 2.34 & & 4.36 \\
   %\makecell{AG-LSEC Eigen} & X & X & X & X & X & X & X \\
   \makecell{AG-LSEC Early Fusion} & \textbf{1.56} & \textbf{1.58} & \textbf{1.56} & \textbf{2.86} & \textbf{2.48} & \textbf{3.01} \\\hline
 \end{tabular}
 \vspace{-0.5\baselineskip}
\end{table*}

\subsection{Acoustic Speaker Scores}
\subsubsection{End-to-end Neural Diarization}
\label{EEND}

We choose the End-to-end neural diarization (EEND) for our SD system due to its ability to handle overlapping speech efficiently compared to traditional embedding based approaches. In addition, EEND is a natural choice also because it directly provides soft speaker scores that can be used to ground lexical error correction. Given frame-level acoustic features $\{\boldsymbol{x}_t\}_{t=1}^T, \boldsymbol{x_i} \in \mathbb{R}^{1\times F}$, where $t\in\{1,\dots,T\}$ is the frame index and $F$ is the feature dimension, EEND estimates the corresponding speaker label sequence $\{\boldsymbol{y}_t\}_{t=1}^T$ where $\boldsymbol{y}_t = \{y_{s,t}\}_{s=1}^S$ denotes speech activities of $S$ speakers at frame $t$ and is defined as
\begin{align}
    y_{s,t}=\begin{cases}
    0&\text{(Speaker $s$ is inactive at $t$)}\\
    1&\text{(Speaker $s$ is active at $t$)}\\
    \end{cases}
\end{align}
Given the sequence of frame-level acoustic features $\{\boldsymbol{x}_t\}_{t=1}^T$, EEND models speaker diarization as a multi-label classification problem using a neural network $f_\mathsf{EEND}$ as
\begin{equation}
    (\boldsymbol{p}_1,\dots,\boldsymbol{p}_T)=f_\mathsf{EEND}(\boldsymbol{x}_1,\dots,\boldsymbol{x}_T),
\end{equation}
where $\boldsymbol{p}_t =\{p_{s,t}\}_{s=1}^S$ are the frame level posterior probabilities of $S$ speakers at frame index $t$.
The EEND system is trained to minimize the permutation-free loss between the output posteriors  $\boldsymbol{p}_t$ and the reference
speaker label $\boldsymbol{y}_t$, as follows: 
\begin{equation}
 \mathcal{L}_\text{diar} = \frac{1}{TS} \min_{\phi \in \mathrm{perm}(S)} \sum_t \mathrm{BCE}(\boldsymbol{y}_t^\phi, \boldsymbol{p}_t),
\end{equation}
where $\mathrm{perm}(S)$ is the set of all permutations of the sequence ($1,\dots,S$), $\phi = (\phi_1,\dots,\phi_S)$ is the permuted sequence, $\boldsymbol{y}_t^\phi={[y_{\phi_1,t},\dots,y_{\phi_S,t}]}\in\{0,1\}^S$ is the permuted ground-truth labels using $\phi$, and $\mathrm{BCE}(\cdot,\cdot)$ is the binary cross entropy loss. 

\subsubsection{Posterior post-processing}

The EEND frame posteriors tend to be noisy and can produce unreasonably short speaker turns. So, a median filter is applied to the posteriors to produce the smoothed posteriors $\boldsymbol{\hat{p}_t} =\{\hat{p}_{s,t}\}_{s=1}^S$ as
\begin{equation}
\hat{p}_{s,t} = \mathrm{median}(p_{s,t-M/2}, p_{s,t-M/2+1}, \ldots, p_{s,t+M/2})
\end{equation}
where M is the number of frames over which median is calculated.  The ASR system produces a sequence of words $\{\boldsymbol{W_i}\}_{i=1}^N$ along with a set of word timings which can be converted to EEND frame indices $\{t_{i,\text{start}}\}_{i=1}^N, \{t_{i,\text{end}}\}_{i=1}^N$ which denote the start and end frames corresponding to each word respectively, where $N$ is the number of transcribed words in the speech segment. The frame-level smoothed posteriors $\hat{p}_{s,t}$ are pooled at the word-level to provide word-level aggregated posteriors $\{\boldsymbol{a_i}\}_{i=1}^N, \boldsymbol{a_i}=\{a_{s,i}\}_{s=1}^S$ as
\begin{equation}
a_{s,i} = \frac{1}{t_{i,\text{end}} - t_{i,\text{start}}} \sum_{t=t_{i,\text{start}}}^{t_{i,\text{end}}} \hat{p}_{s,t}
\end{equation}
The word-level posteriors are normalized in the range (0,1) to produce the word-level acoustic speaker scores $\{\boldsymbol{\hat{a}_i}\}_{i=1}^N, \boldsymbol{\hat{a}_i}=\{\hat{a}_{s,i}\}_{s=1}^S$ where
\begin{equation} \label{eq:example}
\hat{a}_{s,i} = \frac{a_{s,i}}{\sum_{s=1}^{S} a_{s,i}}
\end{equation}

\subsection{Acoustic Grounding}
\subsubsection{Early Fusion}
%After the post-processing step, every word $\{\boldsymbol{W_i}\}_{i=1}^N$  has a corresponding set of speaker scores $\{\boldsymbol{\hat{a}_i}\}_{i=1}^N$. 
%We train and infer the AG-LSEC on windows containing a maximum of K distinct speakers locally in a window, where K<=M. So, the acoustic features in every window are sub-sampled to extract features of dimension K. This is done by sampling the largest K acoustic features for every word. 
%We do not perform Error Correction if K>M, that is if the number of globally detected speaker clusters are lesser than the number of speaker the AG-LSEC is trained for. 
The words \(W_i\) are tokenized into sub-words and passed to the backbone LM to obtain contextual sub-word embeddings \(\{\boldsymbol{E_i}\}_{i=1}^{N_1}, \boldsymbol{E_j} \in \mathbb{R}^{1\times W}\) where \(\boldsymbol{N_1}\) is the number of sub-words in the word sequence and W is the word embedding dimension. 
\begin{equation} \tag{8}
\{\boldsymbol{E_i}\}_{i=1}^{N_1}=EncoderLM(\{\boldsymbol{W_i}\}_{i=1}^N)
\end{equation}
The word level speaker scores \(\boldsymbol{\hat{a}_{i}}\) are mapped to the first sub-word token of the word if the word has more than 1 token and a special “don’t care” vector is assigned to any of the subsequent word tokens. The corrected word-level speaker posteriors  \(\{\boldsymbol{z_{i}}\}_{i=1}^N, \boldsymbol{z_i} \in \mathbb{R}^{1\times M}\) are then extracted as
%sub-word level embeddings \(E_j\) are concatenated with \(\boldsymbol{\hat{a}_{i}}\) or \(\boldsymbol{\hat{f}_{i}}\) to form the audio grounded lexical features for the Front-end Transformer Encoder as shown in Figure 1c.
\begin{equation} \tag{9}
\{\boldsymbol{c_i}\}_{i=1}^{N_1}=concat(\{\boldsymbol{E_i}\}_{i=1}^{N_1}, \{\boldsymbol{\hat{a_i}}\}_{i=1}^{N_1})
\end{equation}
\begin{equation} \tag{10}
\{\boldsymbol{z_{i}}\}_{i=1}^N=Softmax(Encoder(\{\boldsymbol{c_i}\}_{i=1}^{N_1}))
\end{equation}
where posteriors corresponding to the first sub-word tokens are selected to represent the word level speaker posteriors. These posteriors are optimized for Cross Entropy (CE) loss on the ground-truth speaker labels. This early fusion process is shown in Figure 1c.
 
\subsubsection{Late Fusion}
%The word level speaker labels $\{\boldsymbol{s_i}\}_{i=1}^N, \boldsymbol{s_i} \in \mathbb{R}$ for the LSEC model are obtained by assigning the speaker with the largest word-level acoustic scores $\{\hat{a}_{s,i}\}_{s=1}^S$
%\begin{equation} \tag{11}
%s_{i} = \argmax_{s} (\hat{a}_{s,i})
%\end{equation}
The word level speaker labels for the LSEC model are obtained by assigning the speaker with the highest word-level acoustic scores from (\ref{eq:example}) for each word. For the acoustic grounding, we perform lexical correction with the LSEC model and fuse the word level acoustic speaker scores, $\{\boldsymbol{\hat{a}_i}\}_{i=1}^N$ with the corresponding word level lexical posteriors from the LSEC model. A simple Neural network (FusionNet) takes in the acoustic and lexical scores and predicts the correct speaker label as shown in Figure 1a. 
%One advantage of the late fusion mechanism is the relatively low paired audio-text data requirements since the FusionNet can be a lightweight model with low data requirements. 
 
\subsection{Training Methodology}
The AG-LSEC model is trained by initializing it with the LSEC model (impact of the initialization is shown in Figure 2) since the speaker scores used in the AG-LSEC correspond to the soft scores of the speaker labels used to train the LSEC model. It is trained with paired audio-text data, where the audio is diarized with the SD system, transcribed by an ASR system and reconciled to get the words and the word level acoustic scores as in Figure 1.

Similar to the AG-LSEC training data, word-level speaker scores are extracted using the paired audio-text data and the lexical posteriors are extracted using the LSEC model which are fed to the Fusion Net to predict the corrected speaker label. 

The AG-LSEC, LSEC and Fusion Net are all trained over local windows with two hypothesized speakers.
\subsection{Inference Setup}
During inference, we perform error correction on overlapping sliding windows with a fixed number of ASR transcribed words. Similar to LSEC \cite{paturi23_interspeech}, AG-LSEC can also handle cases where more than two speakers are detected globally across the audio by only correcting sliding windows comprising of one or two speakers and bypassing the remaining windows.

\section{Experiments}
\label{exps}

\subsection{Data and Metrics}
We use the Fisher dataset \cite{40LDC, 41LDC} to train the AG-LSEC system. We create a single channel version of Fisher by merging both the channels and leverage the transcripts for each channel along with the timestamps as the ground-truths similar to \cite{paturi23_interspeech, wang2024diarizationlm} and we use the same held-out Fisher test split for evaluation as used in \cite{wang2022highly, paturi23_interspeech, wang2024diarizationlm}. For evaluation, in addition to the held out Fisher test split, we use 2 other telephony datasets, namely CALLHOME American English (CHAE) \cite{39LDC} and RT03-CTS \cite{42LDC}. %which are primarily 2-speaker calls.
%We also evaluate on the two-speaker only set of CHAE, the CH-109 dataset \cite{cyrta2017speaker} by fixing the number of clusters to 2 as well as automatically determining the number of speakers in the 1st pass SD system.

The metric we use to evaluate our model is the Word Diarization Error Rate (WDER) proposed in \cite{shafey2019joint} as it aptly captures both ASR and SD errors at the word level and accounts for errors in the overlapping speech regions as well. In order to align the multi-speaker references to multi-speaker hypothesis including regions of speech overlap, we use asclite tool \cite{fiscus2006multiple}.

\subsection{Baseline System}

The baseline EEND system follows \cite{fujita2019end} and consists of 6 stacked self-attention-based Transformer layers, 8 attention heads with a hidden size of 256 and 1024 internal units in the position-wise feed-forward layer. It is trained on 3-speaker simulated mixtures similar to \cite{fujita2019end2} and adapted on the CALLHOME (CH) dataset \cite{doddington2000nist}. During inference, a median filter is applied over 11 frames in the post-processing of EEND posteriors and the entire audio is diarized in a single pass without the use of any segmentation. This baseline EEND
system achieves a DER of 7.36 on the CH test set compared to a DER of 10.76 reported in \cite{fujita2019end} . 

%The baseline EEND system is comparable to state-of-the-art diarization systems across telephony datasets and achieves a a DER of X on CHAE \cite{39LDC} which is a stronger baseline than the one Y reported in []. 
We use an end-to-end ASR system \cite{zhou2020rwth,yang2022conformer} with a Conformer Acoustic model \cite{gulati2020conformer} and a n-gram Language model trained on several tens of thousands of audio and text data. The ASR system %is comparable to state-of-the-art ASR systems across several datasets and 
achieves a WER of 12.2 on the single channel Fisher test set compared to 15.48 reported in \cite{wang2024diarizationlm}. The combined SD-ASR system achieves a WDER of 2.56 on Fisher test set which is a much stronger baseline compared to the WDER of 5.32 reported in \cite{wang2024diarizationlm}.
%The baseline EEND-ASR system is comparable to a state-of-the-art speech transcription system across telephony datasets and achieves a WDER of  on CHAE \cite{39LDC} compared to Y reported in []

\subsection{AG-LSEC System}
We follow the same architecture and training procedure for the LSEC model as outlined in \cite{paturi23_interspeech} which uses the Roberta-base model \cite{liu2019roberta} as the backbone LM and a 1 layer Transformer Encoder of size 128 hidden states for the Front-end model. We use this same model for our late fusion experiments. The backbone and front-end model architectures are the same for the AG-LSEC model as well. The AG-LSEC model is trained with Adam Optimizer with a batch size of 32 and an average sequence length of 30 words per batch. We use a learning rate of 1e-4 and train the model for 20 epochs on a machine with 8 GPUs.

\subsection{Results}

We benchmark WDER on the test sets with our proposed acoustic grounding techniques and compare these to the baseline ASR, SD system as well as the LSEC model in Table 1. It can be observed that the proposed AG-LSEC model with early fusion achieves significant relative WDER improvements of 25\% to 40\% across all the test sets and also improves the WDER by more than 15\% relative over the LSEC model. The late fusion model also improves the WDER over the LSEC model, though produces smaller relative gains in the order of of 5-10\%.
\begin{figure}[t]
  \centering
  \includegraphics[width=0.8\linewidth]{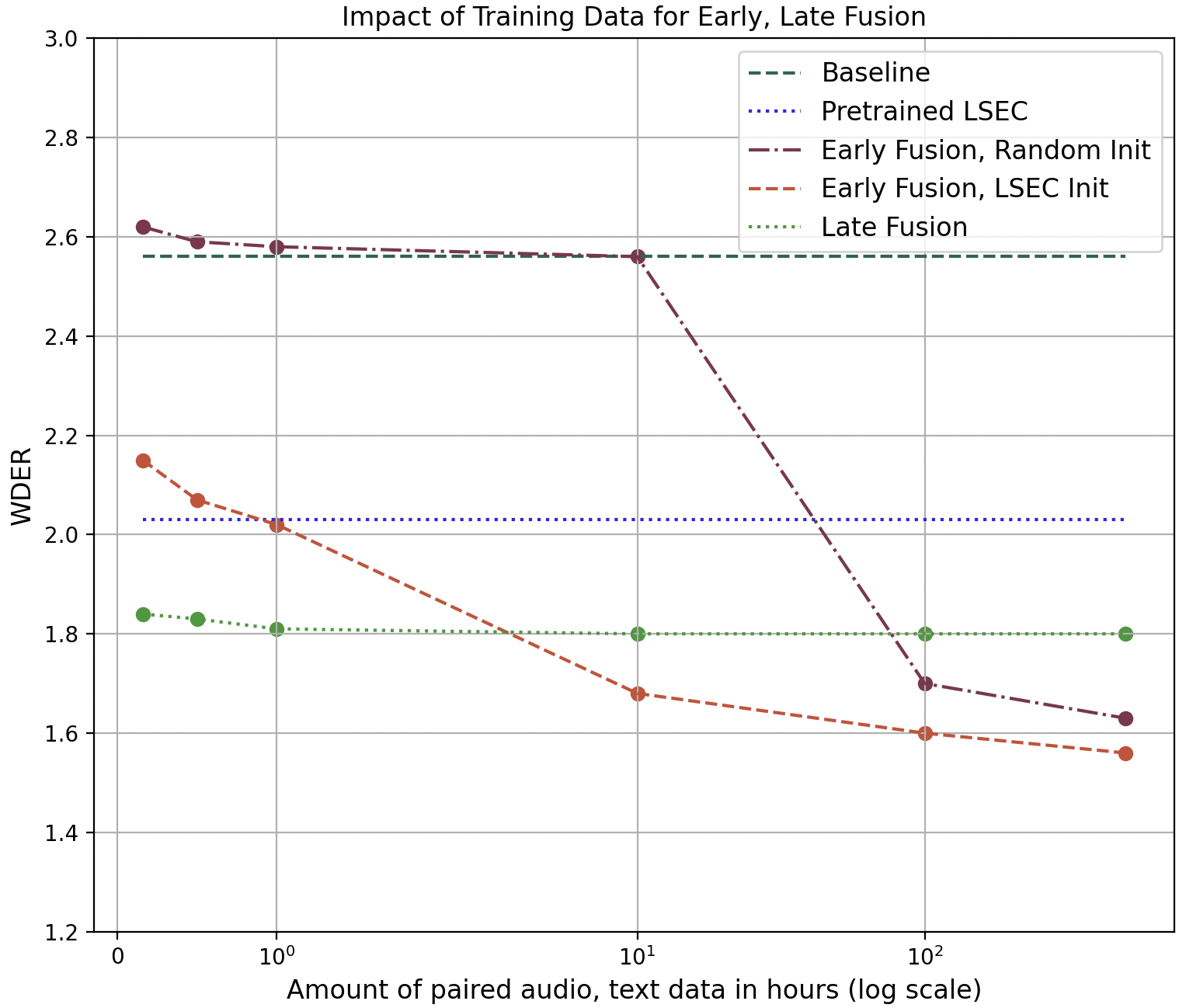}
  \caption{Ablation of the AG-LSEC models with different amounts of training data and initializations.}
  \vspace{-1\baselineskip}
\end{figure}

In order to analyze the paired data requirements for training the models, we evaluate the WDER of the models on Fisher test set when trained with limited amount of speaker tagged, paired audio-text data. For this analysis specifically, we train the LSEC model only on text data with synthetic simulated errors \cite{paturi23_interspeech} and do not include any paired audio-text data. From Figure 2, it can be observed that the late fusion model is able to achieve the same consistent improvement of ~7\% over the text only trained LSEC model with as little as 20 minutes of paired data. It can also be observed that the LSEC initialization for the AG-LSEC model surpasses the late fusion model around 10hrs of data while also much better than the random initialized AG-LSEC model till almost 100hrs of training data, while still being marginally better beyond 100hrs of data. This shows the importance of the LSEC initialization in both cutting down the data requirements as well to achieve better WDER improvements. 

We further analyze the behaviour of the AG-LSEC models in correcting errors whilst also introducing new errors over the baseline SD-ASR system. For this, we benchmark the LSEC model, AG-LSEC model with Early and Late Fusion. These results are documented in Table 2. It can be seen that the AG-LSEC Early Fusion introduces lesser errors than the LSEC model, thus preventing over-corrections. At the same time, both Early and Late Fusion correct more speaker errors than the LSEC model, which also suffers from under-corrections due to its excessive reliance on lexical cues. 

Some qualitative examples of such over-corrections and under-corrections by the LSEC model and its rectification by the AG-LSEC model are presented in Figure 3. It can be seen from 3a) that the LSEC model over-corrects to a lexically more plausible yet incorrect speaker transcript which the audio grounding prevents with the AG-LSEC model. Similarly, from 3b), it can be seen that the LSEC model fails to correct the lexically plausible baseline due to its excessive dependence on lexical characteristics. On the contrary, the AG-LSEC model has the potential to correct such errors due to the inclusion of acoustic speaker scores along with the lexical features, thus improving the correction performance significantly.
\section{Conclusion and Future Work}
In this work, we propose a novel acoustic fusion strategy to the recently proposed Lexical Speaker Error Corrector (LSEC) to correct word-level speaker label errors from a conventional multi-talker speech transcription system. We extract acoustic speaker scores directly from the existing EEND pipeline and outline two fusion strategies by leveraging these speaker scores, both showing improvements over the baseline LSEC system. The early fusion AG-LSEC model achieves significant relative WDER improvements of above 25\% on the test datasets over the the baseline SD-ASR system and also handily beats the LSEC model with relative WDER improvements of over 15\%. %Though the WDER improvements of the late fusion AG-LSEC over LSEC are lesser, the late fusion model is able to achieve these gains with as less than 20 minutes of paired audio-text speaker data.

%As a next step, we plan to integrate the AG-LSEC EF model with EEND-Vector Clustering (EEND-VC) \cite{kinoshita2021integrating, bredin23_interspeech} based SD approaches which have shown significant improvements in diarizing overlapped speech over the traditional SD approaches. The AG-LSEC model can naturally leverage the local speaker posteriors form an EEND model to perform overlap-aware Speaker Error Correction. 
In future, we will experiment with a multi-lingual AG-LSEC model to handle languages other than English and will handle more than 2 speakers locally to increase the coverage of lexical correction.
%In addition, we also plan to also explore leveraging large generative models to synthesize conversational transcripts across multiple domains using curated prompts \cite{chen2023places} to reduce the amount of paired audio-text data required to achieve these improvements.
\begin{table}[t]
  \caption{Percentage of errors corrected and new errors introduced with LSEC and variants of AG-LSEC on Fisher test set.}
  \vspace{-0.2\baselineskip}
  \label{tab:Early Fusion Strategies}
  \centering
    \begin{tabular}{P{30mm}P{20mm}P{20mm}N} 
    \hline\hline
        \multirow{2}{*}{Model}&\multirow{2}{*}{\makecell{Errors\\Corrected (\%)$\uparrow$}}&\multirow{2}{*}{\makecell{Errors\\Introduced (\%)$\downarrow$}} \\ && \\ \hline\hline
        %Model & Errors Corrected (\%)$\uparrow$ &  Errors Introduced (\%)$\downarrow$  \\ \hline\hline
        LSEC \cite{paturi23_interspeech}  & 29.2 & 8.4 \\
        AG-LSEC Late Fusion  & 38.18 & 9.2 \\
        AG-LSEC Early Fusion & \textbf{44.53} & \textbf{6.6}  \\ \hline
    \end{tabular}
    % \vspace{-0.5\baselineskip}
\end{table}
\begin{figure}[t]
  \centering
  \includegraphics[width=1\linewidth]{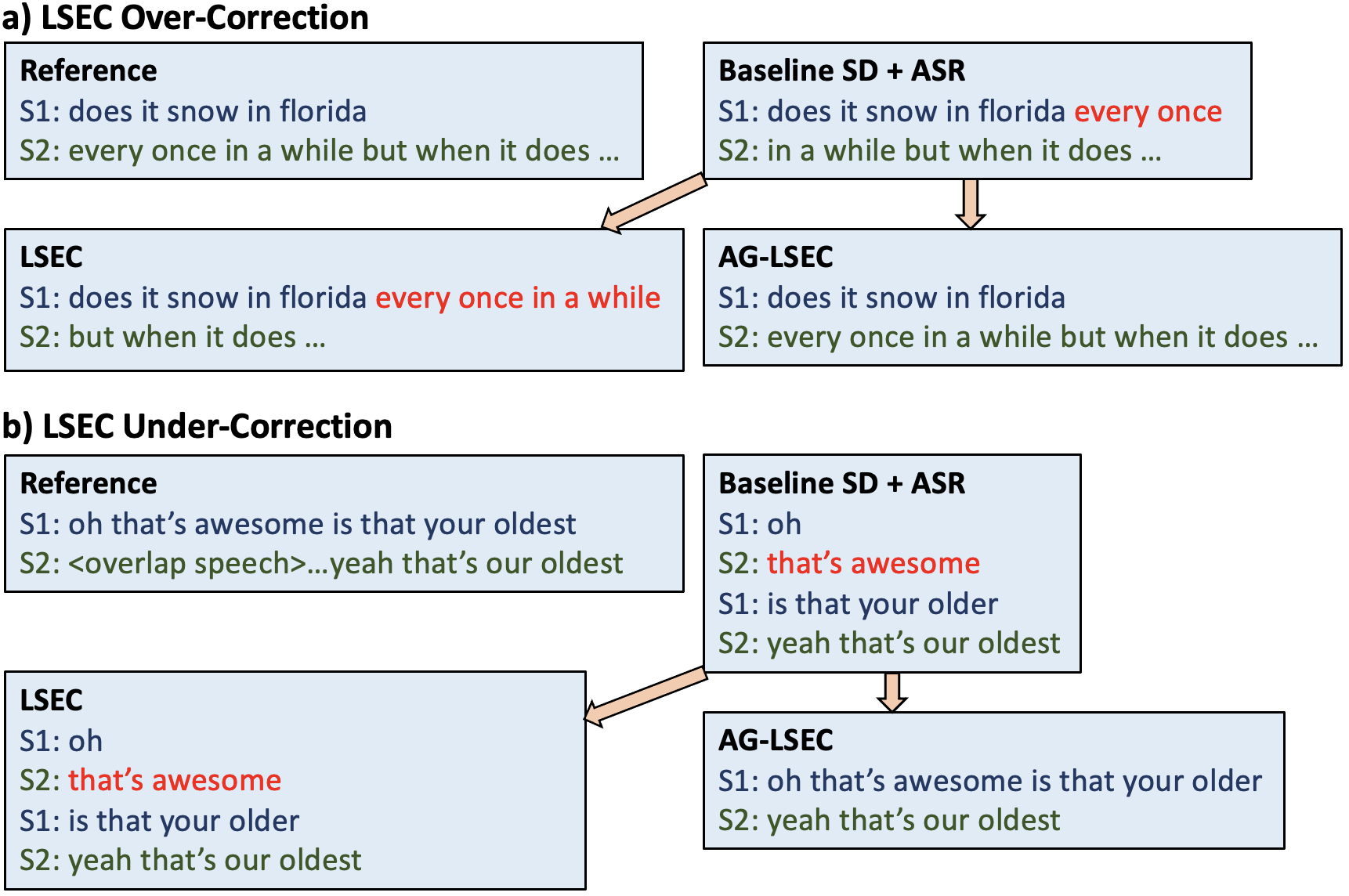}
  \caption{Qualitative examples of over-correction and under-correction with LSEC model rectified by the AG-LSEC Early Fusion model. The words highlighted in red are speaker errors.}
  \vspace{-1\baselineskip}
\end{figure}
\renewcommand{\bibsection}{\section{REFERENCES}}
\bibliographystyle{ieeetr}
{\eightpt
\bibliography{citations}}

\end{document}